\documentclass[fleqn,usenatbib]{mnras}

\usepackage{newtxtext,newtxmath}

\usepackage[T1]{fontenc}
\usepackage{booktabs}
\DeclareRobustCommand{\VAN}[3]{#2}
\let\VANthebibliography\thebibliography
\def\thebibliography{\DeclareRobustCommand{\VAN}[3]{##3}\VANthebibliography}

\usepackage{xspace}
\usepackage{enumitem}
\usepackage{orcidlink}
\usepackage{gensymb}
\usepackage{graphicx}
\usepackage{float}
\usepackage{caption}
\usepackage{subcaption}
\usepackage{soul}
\usepackage{arydshln}

\usepackage{lineno}

\graphicspath{{./figs/}}

\newcommand{\src}{{J1727}} 
\newcommand{\srclong}{{Swift J1727.8-1613}}

\newcommand{\xsho}{\textit{X-Shooter}\xspace}

\title[Optical outburst evolution BHXRB Swift J1727]{Optical outburst evolution of the transient black hole X-ray binary {\it Swift}\,J1727.8$-$1613: Disc response to jet ejections and late-outburst emergence of powerful disc winds}

\author[N. Castro Segura et al.]{N.~Castro~Segura$^{1}$$\orcidlink{0000-0002-5870-0443}$\thanks{E-mail: N.Castro-Segura@soton.ac.uk},
K.~Solomons$^{2,3}$$\orcidlink{0000-0003-2323-5067}$,
J.~M.~Corral-Santana$^{4}$$\orcidlink{0000-0003-1038-9104}$,
C.~Knigge$^{5}$$\orcidlink{0000-0002-1116-2553}$,
P.~A.~Charles$^{5,6,7}$$\orcidlink{0000-0002-4672-7278}$,
\newauthor
M.~Brigitte$^{8,9}$$\orcidlink{0009-0004-1197-5935}$, 
S.~Fijma$^{10}$$\orcidlink{0000-0002-3420-3522}$,
M.~Diaz-Trigo$^{10}$$\orcidlink{0000-0001-7796-4279}$,
A.~G\'urpide$^{5}$$\orcidlink{0000-0002-2256-2704}$,
D.~A.~H.~Buckley$^{3}$$\orcidlink{0000-0002-7004-9956}$,
F.~Carotenuto$^{11}$$\orcidlink{0000-0002-0426-3276}$,
\newauthor
A.~J.~Castro-Tirado$^{12}$$\orcidlink{0000-0003-2999-3563}$,
D.~L.~Coppejans$^{1}$$\orcidlink{0000-0001-5126-6237}$,
M.~Georganti$^{5}$$\orcidlink{0000-0002-3776-9652}$,
A.~Hughes$^{7}$$\orcidlink{0000-0003-0764-0687}$,
K.~S.~Long$^{13,14}$$\orcidlink{0000-0002-4134-864X}$,
J.~Matthews$^{7}$$\orcidlink{0000-0002-3493-7737}$,
\newauthor
I.~Monageng$^{2,3}$$\orcidlink{0000-0002-4754-3526}$,
I.~Pelisoli$^{1}$$\orcidlink{0000-0003-4615-6556}$,
T.~D.~Russell$^{15}$$\orcidlink{0000-0002-7930-2276}$,
D.~Steeghs$^{1}$$\orcidlink{0000-0003-0771-4746}$,
J.~Svoboda$^{8}$$\orcidlink{0000-0003-2931-0742}$,
A.~J.~Tetarenko$^{16}$$\orcidlink{0000-0003-3906-4354}$,
\newauthor
F.~M.~Vincentelli$^{17}$$\orcidlink{0000-0002-1481-1870}$,
A.~G.~W.~Wallis$^{5}$$\orcidlink{0000-0003-0770-9015}$
\\
Affiliations are listed at the end of the paper
}

\date{Accepted XXX. Received YYY; in original form ZZZ}

\pubyear{2026}

\begin{document}
\label{firstpage}
\pagerange{\pageref{firstpage}--\pageref{lastpage}}
\maketitle

\begin{abstract}
Swift J1727.8$-$1613 is a newly discovered transient low-mass X-ray binary harbouring a stellar-mass ($\sim 10M_\odot$) black hole. We present state-resolved VLT/X-Shooter optical spectroscopy of its 2023 outburst, sampling the luminous hard-to-soft and late soft-to-hard transitions. 
During the onset of the brightest radio flare, He\,\textsc{ii} flux rises relative to adjacent epochs, with reduced peak-to-peak separation and full-width-half-maximum, consistent with enhanced irradiation shifting line emissivity to larger radii. We detect no contemporaneous change in the line base tracing the inner disc. 
The most dramatic change occurs at the onset of the dim-hard state, when strong, broad (higher-order) Balmer lines appear in absorption, and He\,\textsc{ii} remains in emission, but becomes highly asymmetric. 
While the hardening of the X-ray spectrum likely promotes the reappearance of an underlying disc photosphere, the kinematic alignment between the Balmer absorption ($v_w\sim-750\,\mathrm{km\,s^{-1}}$) and the suppressed blue peak of He\,\textsc{ii} suggests a unified origin in a massive, cool ($T\lesssim10^{4}\,\mathrm{K}$) accretion disc wind. 
Radiative transfer simulations demonstrate that such asymmetric He\,\textsc{ii} profiles are naturally produced in a rotating and accelerating outflow.
Using the Sobolev approximation, we estimate the wind mass-loss rate to be $\dot{M}_w\gtrsim10^{-9}\,M_\odot\,\mathrm{yr^{-1}}$, comparable to the instantaneous accretion rate and a significant fraction of the secular mass-transfer rate from the donor. If persistent at quiescent-level X-ray luminosities, this outflow could strongly impact the system's secular evolution.
\end{abstract}




\section{Introduction}
Low-mass X-ray binaries (LMXBs) are interacting systems harbouring either a black hole (BH) or a neutron star accreting matter from a low-mass companion star ($\lesssim1 \mathrm{M}_{\odot}$), usually via Roche-lobe overflow. Due to angular momentum conservation, the infalling material tends to form an accretion disc, where gravitational potential energy is dissipated in the form of radiation \citep{ShakuraSunyaev1973}. 
Transient LMXBs undergo sporadic outburst episodes produced by thermal–viscous instabilities in the accretion disc \citep{Lasota2001NewAR..45..449L,McClintock2006}. 
These outbursts are understood as a sudden increase in the mass accretion rate, during which systems brighten by several orders of magnitude at all wavelengths, making them some of the brightest astronomical objects in X-rays \citep{FrankKingRaine2002apa..book.....F}.

Transient black hole X-ray binaries (BHXBs) exhibit distinct spectral states in X-rays, most notably the so-called hard- and soft-states, during which their energy output is dominated by hard and soft X-rays, respectively. 
Over the course of an outburst, these systems follow a hysteresis cycle between these two spectral states. 
During quiescence and on the rise, these systems exhibit a hard X-ray spectrum which is thought to be associated with an optically thin, geometrically thick accretion flow. Near the outburst peak, the inner flow is thought to be replaced by an optically thick, geometrically thin accretion disc. This gives rise to a dominant soft, thermal component until the system returns to the (low-luminosity) hard-state towards the end of the eruption \citep[e.g. ][]{Esin1997ApJ...489..865E,Done2007AARv..15....1D}. 
This hysteresis cycle is clearly seen in the X-ray hardness intensity diagram \citep[HID e.g. ][]{Fender_2004}, where the erupting systems trace a ``q-shaped'' path.

Traditionally, these two spectral states have been associated with two types of outflows. 
During the hard state, strong radio emission is present, thought to be associated with a steady compact radio jet \citep[e.g. ][]{Fender_2004}. 
When the source transitions to the soft state, the compact-jet emission is quenched and discrete ejecta (“blobs”) are launched from the system, being detected in the radio band, and occasionally, even X-ray wavelengths \citep{MirabelRodriguez:1994Natur.371...46M,Espinasse:2020ApJ...895L..31E,wood2025}. 
On the other hand, X-ray absorption lines produced by disc winds are only seen in the soft-state \citep[in sufficiently high inclination systems;][]{Ponti2012MNRAS.422L..11P, Parra:2024A&A...681A..49P}. 
The apparent lack of such signatures during hard states was long suspected to reflect selection effects, potentially because the outflow becomes highly ionized and therefore difficult to detect \citep{Diaz-TrigoBoirin2016AN....337..368D}.

However, over the last decade, disc wind signatures in optical recombination lines have been detected in several systems during hard states \citep{Panizo-Espinar:2022A&A...664A.100P}. 
Contrary to the hot, highly ionised winds observed in X-rays, these cool optical winds require high densities, implying high mass outflow rates that might regulate the outburst evolution \citep{Munoz-Darias2016, Tetarenko2018Natur.554...69T,Casares2019MNRAS.488.1356C}.
The relationship between the X-ray and optical disc wind signatures remains poorly understood. However, recent simultaneous detections at multiple wavelengths suggest that (at least in some instances), they are different phases of the same outflow \citep{Castro-Segura2022Natur.603...52C,MunozDarias_Ponti:2022_Xray_optical}. Additionally, some studies suggest that "cold winds" are always present, with only their ionization state changing slightly during state transitions \citep{SanchezSierras2020AA...640L...3S}.
Taken together, the emerging picture of disc winds is complex, and our understanding of this critical system component remains fragmented. An obvious way to improve this situation is to obtain state-resolved spectroscopy of BHXBs across a broad wavelength range.
 
Here, we present the optical spectroscopic outburst evolution of the BH transient \srclong\ (hereafter \src). On 2023 August 24, \src\ triggered the {\it Swift Burst Altert Telescope} ({\it BAT}), which classified the transient as a gamma-ray burst \citep[GRB 230824A;][]{Page:2023GCN.34537....1P}. Shortly thereafter, the MASTER-Tunka robotic telescope \citep{Lipunov2010} reported a rising optical counterpart \citep{Lipunov2023}.
\citet{Negoro2023} also reported the source's detection by the {\it MAXI/GSC} telescope with a rapidly increasing flux, reaching more than 2\,Crab ($10-20$\,keV) in less than 90\,min, which ruled out an extragalactic origin. The hard and rapid flaring exhibited during the early stages of the outburst resembles that of V404 Cyg \citep[also V4641 Sgr and Swift 1858.4$-$0814; ][]{J1858NusHare2020ApJ...890...57H,Negoro2023}. The transient quickly reached V$\simeq 12$ mag in the optical (starting from $\sim 20^\mathrm{th}$ mag in quiescence) and also became the brightest X-ray object in the sky \citep[e.g. ][]{Baglio2023}. Its unprecedented brightness gathered wide attention, triggering follow-up campaigns over all energy bands \citep[e.g. ][]{Hughes2025ApJ...988..109H}. 
These observations have provided estimates of key system parameters, which we summarise below for context.
Here, we use medium-resolution optical spectroscopy gathered from VLT/X-Shooter to track the evolution of the disc and outflow signatures in \src\ across its 2023 outburst and associated state transitions, building on the early report of strong emission and absorption features (ATel~\#16208; \citealt{Castro-Tirado2023}).

\subsection{System parameter constraints}

The dynamical study conducted by \citet[][hereafter MS25]{MataSanchez2025A&A...693A.129M}, as \src\ approached quiescence, revealed a K4V companion star with $T_{\rm eff}\simeq 4600\mathrm{K}$ orbiting the compact object with an orbital period of $10.804\pm0.001\,\mathrm{hr}$.  This yielded a binary mass function $f(M)=2.77\pm0.09\,\mathrm{M_\odot}$, implying that \src\ harbours a black hole primary.

However, knowing the precise BH mass requires knowledge of the orbital inclination, and this remains controversial. X-ray spectroscopy of \src\ suggests relatively low inclinations ($i\sim30$–$50^\circ$; \citealt{Peng_Sw1727:2024ApJ...960L..17P,Svoboda_pol_drop_Sw1727:2024ApJ...966L..35S}), while the strength of type-C QPOs modestly favours higher values \citep{Ma_Sw1727_HXMT:2025MNRAS.543.1748M}. The measured jet inclination implies an upper limit $i<74^\circ$ \citep{Wood_Sw1727:2024ApJ...971L...9W}, while the jet dynamics set a tighter value of $i\lesssim67^\circ$ \citep{Wood:2025ApJ...984L..53W}. The source shows no evidence of dipping behaviour at any outburst stage, suggesting $i\lesssim60^\circ$, which is supported by the absence of ellipsoidal modulation in quiescence, down to an $r$-band dispersion of $\sigma_r=0.14$ (MS25; values consistent with the pre-outburst photometry gathered by ZTF). These values are only slightly higher than the orbital modulation amplitude in the BH binary A0620$-$00 ($i\simeq51^\circ$; \citealt{Cantrell:2010ApJ...710.1127C}).

On the lower end of orbital inclination values, the strongest constraints are obtained from $f(M)$. Given the observed MS25 donor spectral type, keeping the donor mass $<1M_\odot$ requires $i\gtrsim45^\circ$, and also implies a BH mass $M_\bullet\simeq10\,\mathrm{M_\odot}$.

As we show below, our results favour a moderate inclination. Moreover, recent optical polarimetry by \citet{Nitindala:2025arXiv251208716N} suggests a jet–orbit misalignment of $15^\circ$ \citep[a fairly common value in BHXBs; e.g.][]{Miller-Jones:2019Natur.569..374M}. Such a misalignment reconciles the tension with the results from X-ray timing and radio observations, while supporting a moderate inclination ($45^\circ \lesssim i \lesssim 55^\circ$). This, in turn, implies a black hole primary with a mass of $M_\bullet \sim 6.7\,\mathrm{M_\odot}$ for $i\simeq50^\circ$, a value consistent within the bulk of confirmed stellar-mass black holes \citep{Corral-Santana2016,Jonker:2021ApJ...921..131J}. These constraints provide the context for interpreting the state-resolved line evolution presented below.

\section{Data Sets}\label{sec:data sets}

\begin{table}
\centering
\caption{Summary of \src\ X-Shooter observations, divided into two sub-campaigns$^*$.}
\begin{tabular}{cccccccr}
\toprule
\quad \textbf{Epoch} & \textbf{Orbital} & \textbf{MJD} & \textbf{Accretion State} \\
 & \textbf{phase} & &  \\ 
\midrule
\multicolumn{4}{l}{Hard-to-soft (early outburst, bright part of HID Fig~\ref{fig:HID})}\\
\midrule
\quad 1   & 0.76 & 60183.01 & Hard \\
\quad 2   & 0.62 & 60191.05 & Hard \\
\quad 3   & 0.79 & 60201.03 & Hard \\
\quad 4   & 0.40 & 60213.01 & Hard \\
\quad 5   & 0.65 & 60218.98 & Hard \\ \hdashline
\quad 6   & 0.01 & 60224.99 & Soft \\
\quad 7   & 0.39 & 60231.01 & Soft  \\
\quad 8   & 0.04 & 60234.01 & Soft \\
\quad 9   & 0.14 & 60239.00 & Soft \\
\midrule
\multicolumn{4}{l}{Soft-to-hard (towards end of outburst)}\\
\midrule
\quad 10  & 0.75 & 60369.38 & Soft \\
\quad 11  & 0.02 & 60393.36 & Hard  \\
\bottomrule
\end{tabular}
\\
$^*$ dashed line indicates hard to soft state transition as defined by\\ \cite{Bollemeijer2023,Bollemeijer2023a}. 
A more detailed observing log \\ can be found in Table~\ref{tab:xs_obs_full}.
\label{tab:xs_obs}
\end{table}

\subsection{Optical/near-IR spectroscopy}

We obtained 13 mid/high-resolution spectra of \srclong\ during 11 distinct epochs using the \xsho\ spectrograph \citep[][]{XShooter}, mounted on the \textit{Very Large Telescope (VLT)} Unit Telescope 3 (UT3, Melipal), as part of a broader multi-wavelength campaign. \xsho\ offers simultaneous wavelength coverage from 3000\,\AA\ to 2.4\,$\mu m$\ across its three observing arms (UVB, VIS, and NIR).
The observations were divided in two campaigns, before and after the Sun constrain. The first campaign covered the beginning of the outburst from 27th August to 21st October 2023. The second campaign consist of two epochs on the 29th of February and 3rd of March covering the soft-to-hard transition\footnote{Program IDs: 111.265Y \& 112.26W7}. 

Each epoch consisted of multiple exposures using an AABB nodding pattern with the number of exposures on each nod equal to half of the total number of exposures (effectively $N/2\times AB\times N/2$ with $N$ the total number of exposures). A slit width of 1$^{\prime\prime}$/0.9$^{\prime\prime}$/0.9$^{\prime\prime}$ was used across all arms and epochs for UVB/VIS/NIR, respectively. The only exception was Epoch 9b, for which we employed a 1.2$^{\prime\prime}$ slit to account for the large effective seeing due to the low elevation at the time of the observations. 

The data were reduced using the latest available version of the \xsho\  pipeline within the \texttt{EsoReflex}\footnote{\href{https://www.eso.org/sci/software/esoreflex/}{EsoReflex: https://www.eso.org/sci/software/esoreflex/}}\citep{EsoReflex2013A&A...559A..96F} environment, and telluric absorption features were corrected using \textit{molecfit} \footnote{\href{https://www.eso.org/sci/software/pipelines/skytools/molecfit}{Molecfit: https://www.eso.org/sci/software/pipelines/skytools/molecfit}} \citep{SmetteA2015}. 
All wavelengths and timestamps were converted into the barycentric rest frame and corrected for line-of-sight proper motion using the solutions from MS25\footnote{systemic velocity $\gamma = -181 \pm 4$\,km\,s$^{-1}$.}. 
The timestamps were corrected using \texttt{jpl} ephemerides by applying the methods implemented in \texttt{astropy.time} \citep{Astropy2018AJ....156..123A}. 
All data were binned to two pixels per resolution element determined by the resolving power of our instrumental setup. An interstellar extinction correction was applied based on a reddening estimate derived from near-UV spectroscopy [$E(B-V)=0.37 \pm0.01 (stat) \pm0.025 (sys)$; \citealt{Burridge2025arXiv250206448B}]. 
A summary of the observations, with the corresponding orbital phase of the secondary and spectral state is presented in Table~\ref{tab:xs_obs} and illustrated in Figure~\ref{fig:HID}. A more detailed version of the observing log can be found in Table~\ref{tab:xs_obs_full}.

To test for short-timescale emission line variability \citep[e.g.][]{charles_shortperiod_wind}, we also obtained six high–time-resolution observations of \src\ with the Robert Stobie Spectrograph (RSS) on the Southern African Large Telescope (SALT; PID: 2021-2-LSP-001) in frame-transfer mode (10–12 s exposures). Three runs were taken in the bright hard state (26, 29 Aug; 17 Sep 2023) and three in the low–hard state near outburst end (14, 19, 20 Mar 2024). All used the PG3000 grating (covering $\sim 4200-4900$ \AA), centred on He\textsc{ii}\,$\lambda4686$. However, no significant line profile changes were seen.

\section{Analysis \& Results}

\begin{figure}
    \centering
    \includegraphics[width=0.99\linewidth]{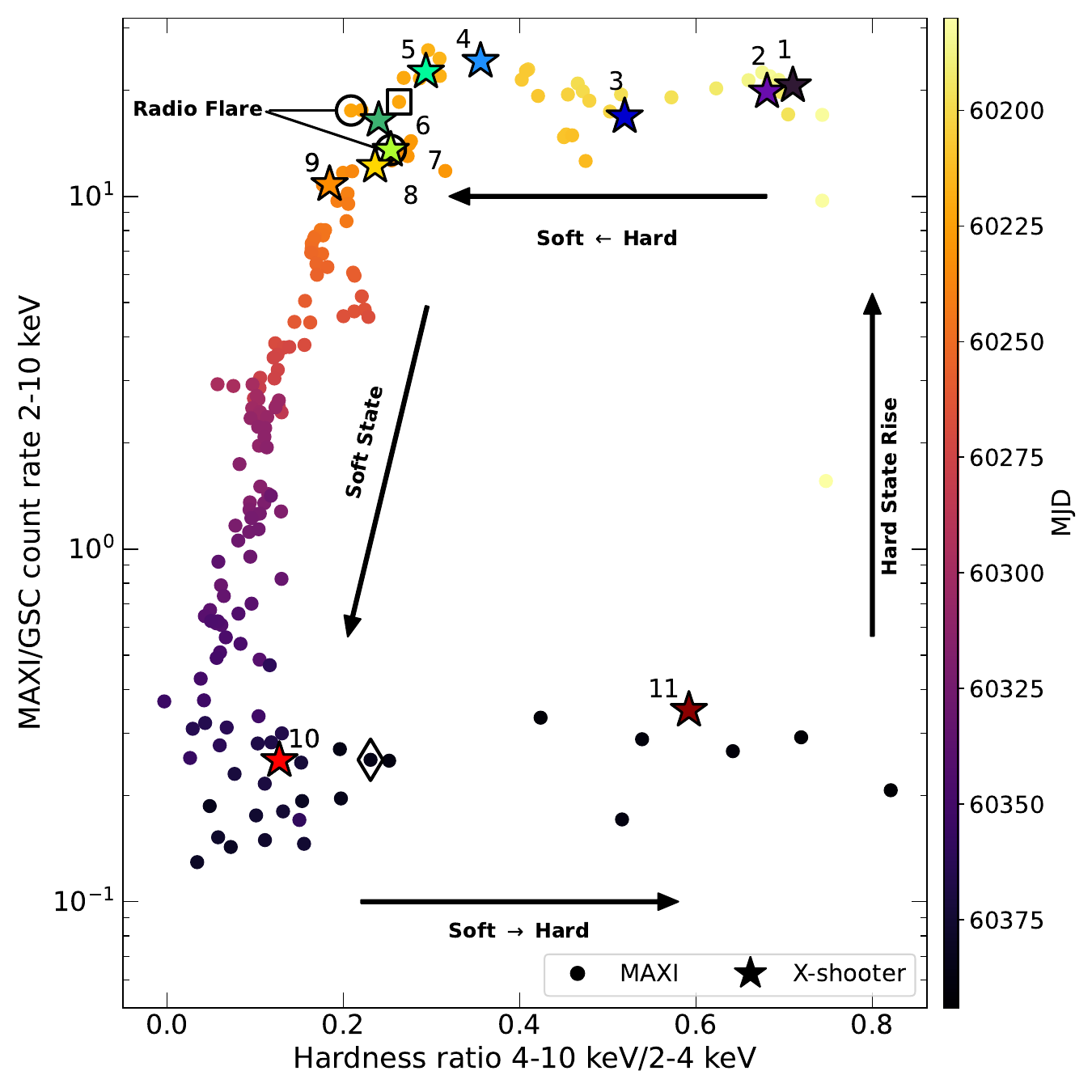}
    \caption{Hardness-intensity diagram of \src\  during its discovery outburst. The 11 spectroscopic epochs gathered with X-Shooter are marked as indicated in the legend. The open circles represent the peak of the two bright radio flares reported by \citet{Hughes2025ApJ...988..109H}. The open square highlights the onset of the hard-to-soft state transition \citep{Bollemeijer2023, Bollemeijer2023a} while the open diamond indicates the onset soft-to-hard state transition \citep{Podgorny20204atel}}
    \label{fig:HID}
\end{figure}

As shown in Figure~\ref{fig:HID}, during its discovery outburst, \src\ followed the canonical q-shaped hysteresis pattern in the hardness–intensity diagram (HID) typical of most transient BH-LMXBs \citep[e.g.,][]{Fender_2004}. In this figure, we present the HID tracing the outburst evolution in the 2–10 keV energy band as observed by the {\it Monitor of All-sky X-ray Image} \citep[{\it MAXI};][]{MAXI:2009PASJ...61..999M}, using data from its {\it Gas Slit Camera} \citep[{\it GSC}; ][]{MAXI_GSC:2002SPIE.4497..173M, 2011PASJ...63S.635S, 2011GSC_PASJ...63S.623M}, retrieved via the {\it MAXI} "on-demand" archive\footnote{\url{http://maxi.riken.jp/mxondem}}. 
In the HID, the data points corresponding to the eleven X-Shooter epochs are marked with numbered, colour-coded stars.
The hard-to-soft and soft-to-hard transitions as defined by \citet{Bollemeijer2023,Bollemeijer2023a}, and \citet{Podgorny20204atel} are also indicated (with an open square and diamond, respectively). As can be seen in this figure, Epochs 1-9 sample the hard-to-soft state transition, which occurs shortly after Epoch 5. The source entered the day side shortly after Epoch 9, in October 2023. 
Two more visits were obtained when \src\ emerged from the Sun's glare at an X-ray luminosity approximately $100$ times lower than that of the previous epochs. The first one (Epoch 10), was observed during the soft state, the last visit (Epoch 11), was obtained towards the end of the soft-to-hard transition (Sec. \ref{sec: STH results}). Epochs 9 and 10 are both averages of two consecutive observations.

\subsection{Spectroscopic evolution} \label{sec: results}

\begin{figure*}
    \centering
    \includegraphics[width=\linewidth]{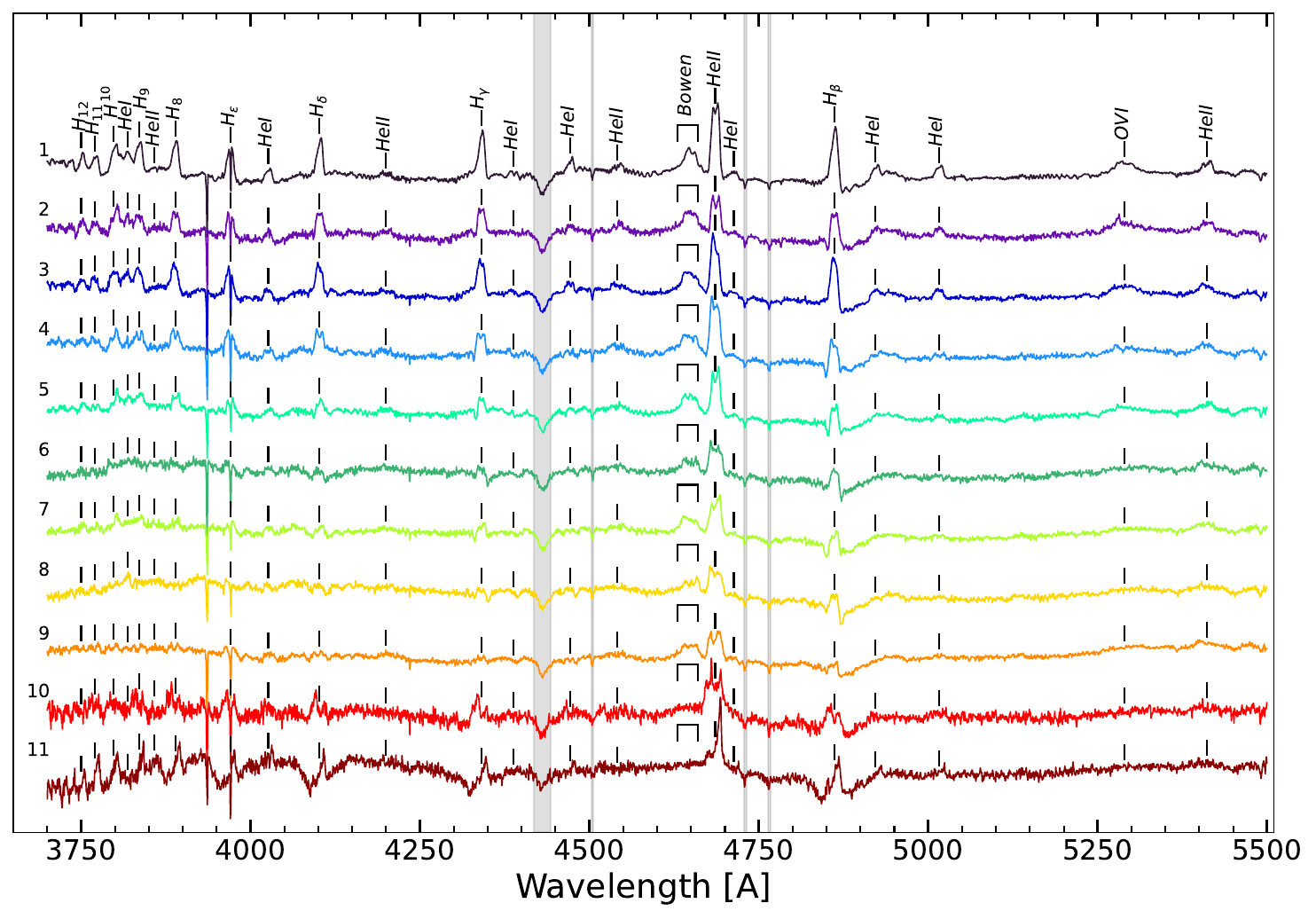}
    \caption{Spectral evolution of \src\ as observed by \xsho's blue-arm. The colour and number of each observation match those in Figure\,\ref{fig:HID}. A vertical offset is applied for clarity. The rest wavelengths of key transitions are marked by vertical ticks. Shaded regions indicate the presence of interestellar lines (except for Ca\,\textsc{ii} H \& K lines at 3933 and 3968 \AA\ respectively).}
    \label{fig: spec evo}
\end{figure*}

\begin{figure*}
    \centering
    \includegraphics[width=0.99\linewidth]{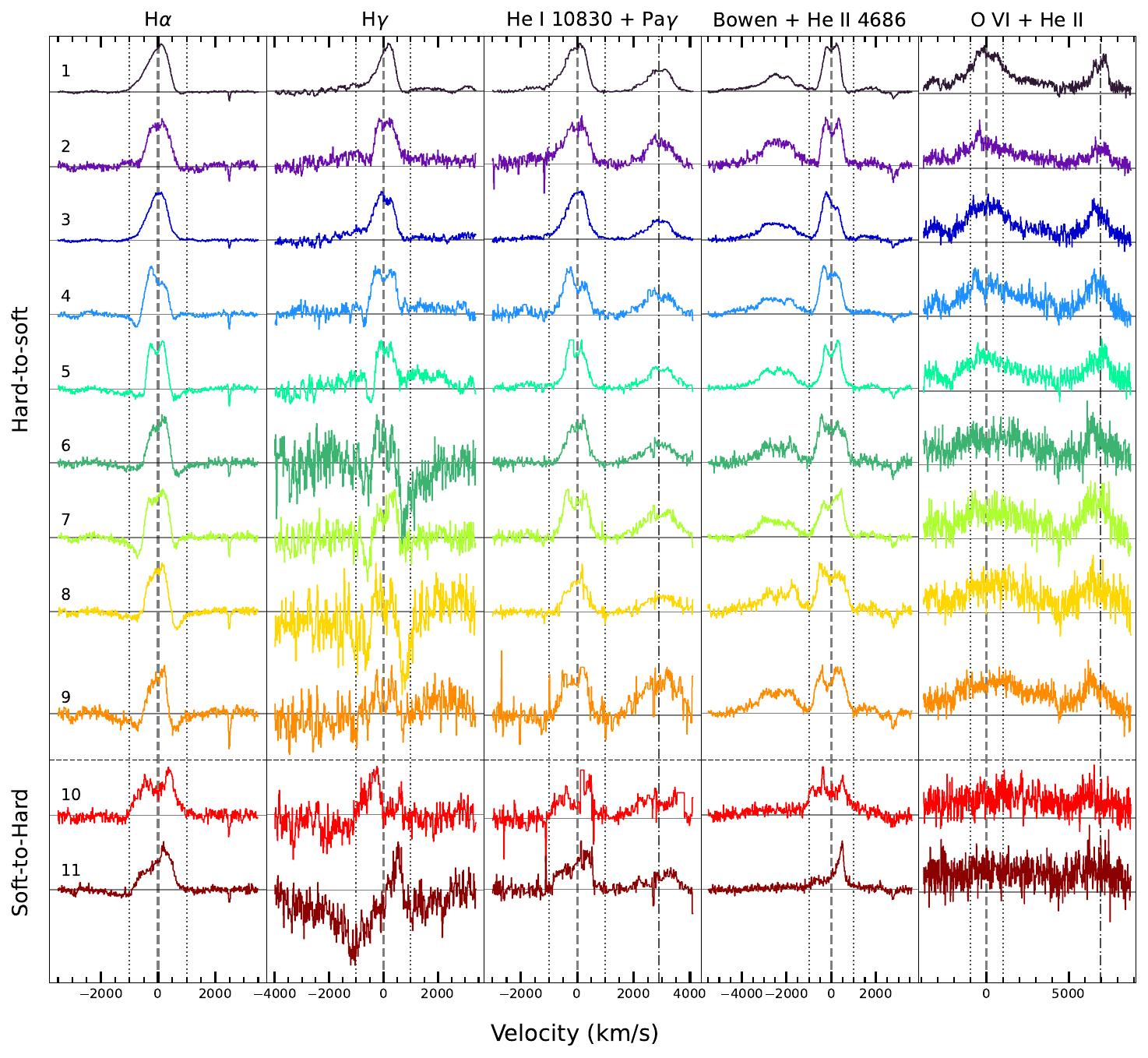}

    \caption{Continuum‐normalized line profiles of \src\ across the 11 \xsho\  observing epochs in velocity space. The figure contain transitions spanning different ionisation potentials, from left to right, the panels show H$\alpha$, H$\delta$, He\,\textsc{i} $\lambda10830$, He\,{\sc ii} $\lambda4686$, and O\,{\sc vi} + He {\sc ii}. Each colour trace corresponds to the same epoch shown in Figure~\ref{fig: spec evo}. Dashed lines indicate zero velocity, while dotted lines mark the approximate positions of transient absorption features on either side of the emission profiles, intended as a qualitative guide to the eye. In the last panel, the zero velocity of O {\sc vi} corresponds to a wavelength of 5284.1\,\AA. Dash-dotted lines in the third and last panels indicate the relative central velocity of Pa$\gamma$ and He {\sc ii}, respectively.
    }
    \label{fig: line evo}
\end{figure*}

In Figure~\ref{fig: spec evo} we show the spectral evolution of the region $\lambda\lambda\approx 3700-5500 $~\AA~across all epochs. This region includes several Balmer and Helium lines, as well as the Bowen blend (at $\lambda \simeq 4650$~\AA). 
To emphasize the gas dynamics, a detailed view of some of the most important lines is shown in Figure~\ref{fig: line evo} in velocity space (i.e. He\,\textsc{ii} $\lambda4686$, He\,\textsc{i} $\lambda10830$, H$\alpha$, H$\gamma$ and O\,{\sc vi}). 

\subsubsection{The bright hard-to-soft state transition} \label{sec: HTS results}
During the early outburst stages, the spectrum of \src\ exhibits both high- and low-ionisation spectral lines, including O\,{\sc vi} and He~{\sc ii}, with a complex continuum around H$\beta$ owing to the presence of strong interstellar absorption.
The first epoch was executed right at the onset of outburst, $\sim 48~\rm h$ after the {\it BAT} alert \citep{Page:2023GCN.34537....1P}. 
This epoch is characterized by the most prominent lines. The peak of the H lines are redshifted, with extended blue wings. 
In contrast, higher ionization lines such as He\,\textsc{ii} $\lambda4686$, exhibit a symmetric double-peaked profile on the top with the body of the line skewed to the blue side. 
All spectra lines during the subsequent epochs before the visibility window closed (i.e. epochs 2--9) typically exhibit a double-peaked emission characteristic of accretion discs \citep{Doppler_tomography}, with a skewed component -- except for Epoch 3 whose line profiles resemble those of the first epoch. 
A detailed view of the He\,\textsc{ii} lines is shown in Fig\,\ref{fig: line evo}, revealing how the width of this double-peaked line (partially controlled by the emissivity distribution), increases as the outburst evolves. This behaviour is consistent with a shift in the emissivity distribution toward smaller radii. While the evolution of peak-to-peak separation suggest the outer-disc radius is moving inwards as the source gets closer to the soft state \citep[][]{Esin1997ApJ...489..865E}. 
However, changes in the irradiation and disc atmosphere structure may also contribute to the observed evolution.
The double-peaked lines are generally asymmetric. The red-to-blue flux ratio shows no correlation with orbital phase, probably indicating the presence of some geometric effect, e.g. warping of the disc, or precession of an elliptical disc \citep{Ogilvie2001MNRAS.320..485O,Haswell2001MNRAS.321..475H}.

Throughout these epochs, the Balmer lines exhibit intermittently appearing absorption features on either (or both) the red and blue wings of the emission line core, starting at about $\pm 1000\,\mathrm{km\,s^{-1}}$. 
These features have previously been reported by \citet{sanchez2024evidenceinflowsoutflowsnearby} and were interpreted as evidence of inflows when the absorption is on the red side, and outflows when the absorption is on the blue side. However, the occasional appearance of absorption on both sides simultaneously may suggest a different origin. 
As shown in the first column of Figure \ref{fig: line evo}, despite changes in depth and detectability, the blue- and red-shifted absorption troughs start at roughly the same velocity on either side, remaining symmetric with respect to the line’s rest position. This hints that the origin of these features may be a symmetric absorption line centred at the rest wavelength filled with emission components, whose properties change with time.  

To test this further, whenever these absorption troughs were present, we fitted the wings of the H$\alpha$ absorption profiles with a broad Gaussian, excluding the emission-core region (as determined by visual inspection).
The fits were carried out using \texttt{lmfit} \citep{newville2025lmfit}, and the associated uncertainties were estimated by bootstrapping each wavelength bin, assuming Gaussian errors as estimated from the adjacent continuum \citep[e.g.][]{Stats_ML_DataMining2014sdmm.book.....I}.
Given the changes observed in the strength of the absorption troughs, not all the epochs are well constrained and some are not constrained at all. 
Despite not being able to fit all epochs, for all the absorption profiles constrained with this method, the measurements are statistically consistent (within $3\sigma$) with the line centre being at the rest position. 
This is true for all epochs where the troughs are present, except for Epoch 7, where we find the absorption centred at $-271\pm46\,\mathrm{km\,s^{-1}}$. Notably, this observation coincides with the onset of the brightest radio flare seen from \src, as discussed below.

\subsubsection{The onset of the soft state \& radio flaring period} \label{sec: radio flaring}

Epochs 6–8 were obtained when the source entered into the soft state. 
The state transition is known for triggering dramatic changes in the accretion and jet geometry, such as (i) the transition from optically thin to optically thick accretion flow as the disc collapses from a geometrically thick to a thin structure \citep[e.g.][]{Esin1997ApJ...489..865E}, (ii) radio flares associated with discrete bipolar jet ejections \citep[e.g.][]{Fender_2004,wood2025}, and (iii) the quenching of the compact jet emission \citep[e.g.][]{Russell2020}. 

\begin{figure}
    \centering
    \includegraphics[width=0.98\linewidth]{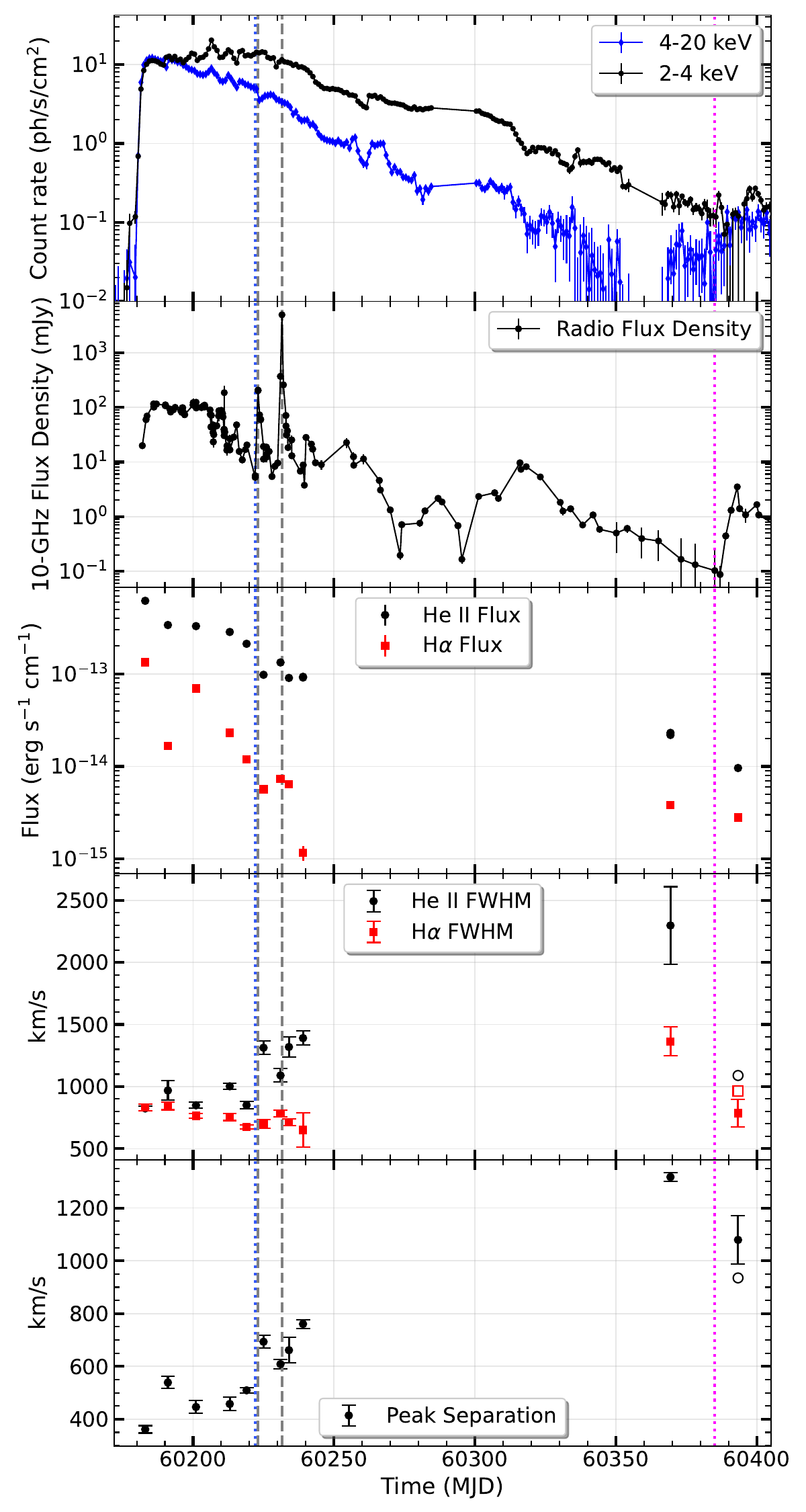}
    \caption{Temporal evolution of X-ray, radio and optical lines through the outburst. From top to bottom, the first pannel shows the X-ray light curve, the second panel shows the $10$ GHz radio light curve from \citet{Hughes2025ApJ...988..109H}, the third and fourth panels shows the Flux and FWHM of He\,\textsc{ii} $\lambda4686$ and H$\alpha$, while the last shows the peak-to-peak separation of the disc-like profiles from He\,\textsc{ii} $\lambda4686$. The vertical dashed lines mark the two radio flares associated with the bipolar ejections observed at the onset of the hard-to-soft transition. Dotted lines indicate the state transitions. We note that the error bars in the flux density are smaller than the symbol size. Given the asymmetric line profiles during the last epoch (see sec. \ref{sec: STH results}), the open symbols are estimated by mirroring the red component about the line centre to represent the missing blue component.
}
    \label{fig: line prop evo}
\end{figure}

To study the co-evolution of the different components in the system, in Fig.~\ref{fig: line prop evo} we compare the X-ray evolution of \src\ with contemporaneous changes in the optical emission lines He\,\textsc{ii} and H$\alpha$. While the X-ray evolution traces the inner accretion region, the optical lines offer a complementary view of the reprocessing and kinematics of the outer accretion flow. 
He\,\textsc{ii} in particular, is an excellent proxy for the ionising continuum in the otherwise unobservable extreme-ultraviolet band \citep[$\simeq55$--$220\,\mathrm{eV}$; e.g.][]{Patterson1985,Marsh1994MNRAS.266..137M}. In our spectra, it shows a clear double-peaked profile across all epochs, consistent with a rotationally dominated disc origin. 
By contrast, the peaks in H$\alpha$ are not reliably distinguished, preventing robust peak-to-peak separation measurements. Moreover, absorption in the line wings (discussed above) renders the flux estimates unreliable. 
Indeed, by looking at the full width at half-maximum (FWHM) of the two lines, it is clear that they do not trace the same evolution, since they exhibit opposite trends (note that for H$\alpha$ we measured it from the continuum level). 
In particular, the H$\alpha$ FWHM decreases as the outburst evolves, contrary to what might be expected from the contraction of the line-forming region. This behaviour instead indicates that the photosphere is extending further out into the disc (c.f. sec. \ref{sec: balmer abs}). We therefore use He\,\textsc{ii} as our primary tracer of the optical line-forming region in the disc.

We characterized the He\,\textsc{ii} profile by measuring its integrated line flux, the peak-to-peak separation, and the FWHM. 
The flux was obtained by integrating the line profiles from the extinction corrected and continuum subtracted spectra.  
The peak-to-peak separation was estimated by isolating each peak and fitting independently using \texttt{lmfit}. We used either a Gaussian model or, alternatively, a second-order polynomial fitted to a handful of pixels around the maximum to determine the centroid of each component. We obtained consistent results with both approaches. The FWHM of the double-peaked profiles was computed numerically by identifying the intersections of the line profile with half of its maximum flux. 

For H$\alpha$, we measured the integrated flux and FWHM using the same approach, but avoiding the absorption troughs. In this case we did not attempt to measure the peak-to-peak separation, because the double-peaked structure is not cleanly resolved. 
In Epoch 10, we excluded the narrow emission component from the fit on the blue side of the rest wavelength, as its centroid is consistent with the donor's orbital solution from MS25, suggesting an origin on either (or both) the irradiated face of the donor and the hotspot. 
In Epoch 11, since the blue wing is clearly absent in He\,\textsc{ii}, for the FWHM and peak-to-peak separation, we also report the values as twice the distance of the red side to the rest position as open symbols.
Uncertainties were estimated using the same bootstrapping method as described above.

As can be seen in Fig.~\ref{fig: line prop evo}, during Epochs 1-5, the He\,\textsc{ii} flux slowly declines as the source softens in X-rays (aside from the first epoch, obtained prior to the outburst maximum). By contrast, the He\,\textsc{ii} peak-to-peak separation and FWHM remain largely constant over this interval. 
The transition to the soft state (Epoch 5$\rightarrow$6), and  the beginning of the radio flaring period, are associated with a drop in the line flux by a factor of $\sim2$, with a small but statistically significant flux increase during Epoch 7. 
The onset of the soft state also produces a sudden increase in both the FWHM and separation of the double-peaked profiles.

The increase in flux during Epoch 7 coincides with the late-rising, near-peak phase of the brightest radio flare reported by \citet{Hughes2025ApJ...988..109H}, occurring $\sim 12$ h prior to the recorded maximum, when the flux had already increased by a factor of $\sim100$ above the pre-flare level. This flare was associated with a jet ejection episode (during which the $1.28\,\mathrm{GHz}$ flux reached $\sim800\,\mathrm{mJy}$). 
At the same time, the He\,\textsc{ii} FWHM and peak-to-peak separation decrease relative to adjacent epochs. If the line profile reflects predominantly the Keplerian motion in the disc, the reduced characteristic velocities and flux increase imply 
additional contribution from larger radii to the line emission during the flare. This is consistent with a transient shift of the dominant He\,\textsc{ii}-emitting region outwards at the time of the jet ejection, suggesting the re-ionization of the outer disc induced by the radio-flare (see Sec. \ref{sec: jet-disc}).

\subsubsection{The soft-to-hard transition} \label{sec: STH results}
 
When \src\ emerged from behind the Sun in February 2024, the source was still active, exhibiting similar hardness-ratios as in the bright soft state, but at a luminosity $\sim100$ times lower (in the $2-10\,\mathrm{keV}$ X-ray band; see Epoch 10 in Fig.\,\ref{fig:HID}). During this time, we carried out two more visits, one in the soft state, and the second towards the end of the soft-to-hard transition (Epochs 10 and 11, respectively). 
As can be seen in Fig.\,\ref{fig: line prop evo}, by Epoch 10 the He\,\textsc{ii} flux has also dropped by more than a factor of 10 with respect to the hard-to-soft transition and about a factor of 5 from the onset of the soft state. 
He\,\textsc{ii} shows a significant increase in both FWHM and peak-to-peak separation, presumably signalling the slow but steady depletion of the ionized accretion disc.

The hypothesis of disc depletion is further supported by the weakening of the Bowen blend relative to He\,\textsc{ii}. Given their respective ionization potentials ($\simeq40\,\mathrm{eV}$ for the Bowen blend and $\simeq 54\,\mathrm{eV}$ for He\,\textsc{ii}), this suggests that the disc's spectral energy distribution is shifted towards shorter wavelengths compared to previous observations, as expected if the outer disc radius contracts \citep[e.g.][]{KimuraDone2019MNRAS.482..626K}. 
In this epoch, we can clearly see double-peaked lines in all the transitions, with the red wing slightly brighter. In particular, He\,\textsc{ii} exhibits a well defined profile with an extra narrow emission component matching the velocity of the donor. This component is associated with the irradiated face of the donor star, and perhaps some additional contribution from the impact point of the accretion stream through the inner Lagrangian point with the rim of the accretion disc (a.k.a. ``the hotspot'').

\begin{figure}
    \centering
    \includegraphics[width=\linewidth]{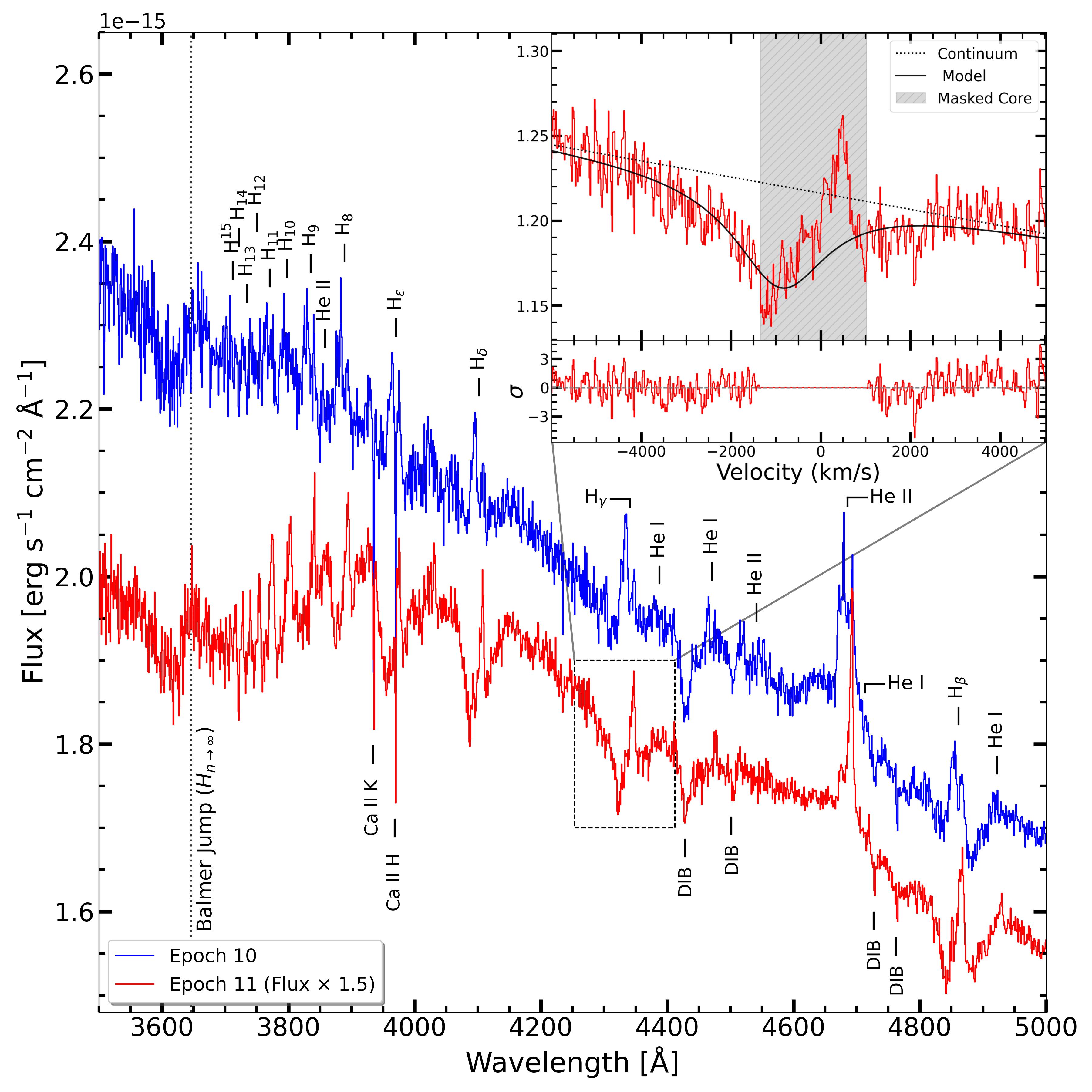}
    \caption{Blue end of the \src\ optical spectrum before he soft-to-hard state transition (Epoch 10) and at the onset of the dim-hard state (Epoch 11). 
    Before the state-transition, during Epoch 10, the spectrum is dominated by emission lines. 
    By Epoch 11, at the onset of the dim-hard state, He~\textsc{ii} is still observed in emission but with a much narrower red-shifted component (See the comparison in Fig.~\ref{fig: epochs10-11}).
    During this last epoch, the overall spectral shape exhibit broad absorption lines (partially filled with narrow emission), somewhat resembling a stellar atmosphere, and a stronger Balmer absorption edge. The broad absorption lines, most notably from $\mathrm{H_\gamma}$ to $\mathrm{H_\zeta}$, are filled with a relatively narrow emission component on the red side of the rest wavelength.
    The contrast in some of the lines allows us to estimate the centroid of the absorption component.
    The inset shows a zoom-in of H$\gamma$ profile, overlaid with the best-fit Lorentzian model obtained by masking the central emission cores (see Table~\ref{tab: abs fit}). 
    These results are consistent with an optically thick component moving in the line sight with a bulk velocity of $v_w\simeq-750\,\mathrm{km\,s^{-1}}$ (see Sections~\ref{sec: STH results} and \ref{sec: outflow discussion} for details). 
    }
    \label{fig: epoch11 only}
\end{figure}

The most striking spectral evolution occurs in Epoch 11, where line profiles change dramatically compared to previous observations (see Fig.~\ref{fig: epoch11 only} and Fig.~\ref{fig: spec evo}). Throughout the outburst, the optical spectrum was dominated by emission lines. However, at the onset of the dim-hard state, the higher-order Balmer lines (most notably H$\gamma$ through H$\zeta$) become dominated by broad absorption features. These profiles are accompanied by relatively narrow emission components situated on the red side of the rest wavelengths. 
Furthermore, Figure~\ref{fig: epoch11 only} also shows a particularly pronounced Balmer absorption edge, consistent with the presence of an optically thick absorbing medium.

These absorption features exhibit a complex, asymmetric structure, with a broad trough that appears distinctly skewed toward the blue. While the observed flux minima are located at approximately $-1000\,\mathrm{km\,s^{-1}}$, the red wings are heavily blended with the emission cores, complicating the determination of the true line centres.
To address this complexity, we explored various modelling configurations, including emission masking (see Fig.~\ref{fig: epoch11 only} inset for an example), multi-component Gaussian modelling, and both single and tied line fits. The resulting centroids vary roughly between $-650\,\mathrm{km\,s^{-1}}$ to $-850\,\mathrm{km\,s^{-1}}$ depending on the approach. For consistency, we report the parameters for the joint Lorentzian fit in Table~\ref{tab: abs fit}. Given the range of results across different modelling approaches, we adopt a representative value of  $v\sim-750\,\mathrm{km\,s^{-1}}$ for the remainder of this work.

He\,\textsc{ii}, on the other hand, is still observed only in emission, although its flux has decreased by a further factor of two compared to the previous (soft-state) observation. Its profile now displays a broad base rising to about a few per cent above continuum, together with a red emission wing that increases smoothly from the rest velocity before dropping sharply at longer wavelengths. The narrow red emission component might naively be associated with the irradiated face of the donor or the hotspot, as suggested for the previous epoch. However, the orbital phase during Epoch 11 was close to zero (Table~\ref{tab:xs_obs}), so that any donor component would be expected near the rest wavelength; this rules out a donor origin for the narrow emission component.
Figure\,\ref{fig: epochs10-11} presents a direct comparison of both Epochs 10 and 11, which shows that the red-shifted peak of He\,\textsc{ii} in Epoch 11 actually coincides with the red peak of the double-peaked emission line from Epoch 10 during the soft state. The broad base of the line follows a similar velocity profile in both epochs.

Moreover, the outflow velocity inferred from the Balmer absorption in Epoch 11 aligns with the blue edge of the broad component underlying the He\,\textsc{ii} emission. Taken together, these characteristics suggest that an underlying double-peaked disc line is still present, but that its blue-shifted side is being suppressed by absorption from the same outflow that is  responsible for the broad Balmer absorption features.

\begin{table}
\centering
\caption{Joint fit of the \src\ H$\gamma$ and H$\delta$ broad absorption profiles in \xsho\ Epoch 11 fitted with Lorentzian profiles. For the multi-component, the emission features were modelled using two Gaussian components, with the line centres and FWHMs of both transitions tied to the same values (see Sec.~\ref{sec: STH results} for details).}
\begin{tabular}{@{}lcr@{}}
\toprule
\textbf{Model} & \textbf{Center (km\,s$^{-1}$)$^*$} & \textbf{FWHM (km\,s$^{-1}$)$^*$} \\
\midrule
\quad Masked Core & $-774 \pm 66$ & $2152 \pm 133$ \\
\quad Multi-component Fit & $-808 \pm 213$ & $1989 \pm 310$ \\
\bottomrule
\end{tabular}
\\
$^*$ Uncertainties represent $1\sigma$ errors from Monte Carlo resampling.\\
\label{tab: abs fit}
\end{table}

\section{Discussion}\label{sec: discussion}

The 2023-2024 outburst of \src, as traced by MAXI, followed the canonical q-shaped trajectory in the HID, transitioning from an initial high-luminosity hard state (epochs 1-5) through a soft state (epochs 6-10), and finally fading into a low-luminosity hard state (final epoch).
In Section~\ref{sec: results} we presented the spectroscopic evolution of the \src\ outburst, as seen by \xsho. In this section, we interpret the main findings from our spectroscopic campaign.

\subsection{The nature of blue- and red-shifted absorption troughs during the bright state.} \label{sec: balmer abs}

One of the most characteristic spectroscopic features throughout the bright state observations are the broad absorption troughs seen on both sides of the emission cores. In some epochs, blue and red absorption are seen simultaneously. These features are most evident in the H$\alpha$ emission wings, but are also visible in other Balmer lines (Fig.\,\ref{fig: line evo}).  
Similar phenomenology has been observed previously in numerous BHXRBs \citep[e.g. GRS $1009-45$, GRO J$1655-40$, XTE J$1118+480$, MAXI J$1807+132$, GRS $1716-249$ and MAXI J$1305-074$;][]{dellaValle_abs_lines:1997A&A...318..179D,Soria_opt_sp_GRO1665:2000ApJ...539..445S, Dubus_opt_sp_XTE1118:2001ApJ...553..307D, Jimenez_ibarra_opt_abs_maxi1807:2019MNRAS.484.2078J,Cuneo_inflows:2020MNRAS.498...25C,Miceli_opt_sp_MAXIJ1305:2024A&A...684A..67M,Corral-Santana:2025A&A...702A.225C}. 

A natural explanation is that the observed profiles arise from the superposition of broad photospheric absorption produced by an optically thick disc, partially filled in by an emission component \citep{Dubus_opt_sp_XTE1118:2001ApJ...553..307D}. Broad Balmer absorption is well known in dwarf novae during outburst and in nova-like variables, where it originates in the optically thick, partially ionised disc photosphere \citep{Beuermann+1990AA...230..326B,Beuermann+1992AA...256..433B}. The emergent spectrum then resembles a stellar atmosphere, although the dissipation in a disc is probably concentrated towards the upper disc layers \citep[c.f.][]{HubenyLong_NL_atms:2021MNRAS.503.5534H}. It is these upper layers and/or the disc wind that can then give rise to the observed emission lines \citep[e.g.][]{Matthews+2015MNRAS.450.3331M,Tomaru_winds_coronal_sources:2023MNRAS.523.3441T}, with additional (typically weaker) contributions from the stream–disc impact region and the irradiated donor face. The visibility of absorption is also dependent on inclination angle: limb darkening can reduce the apparent absorption strength at higher inclinations, yielding profiles dominated by pure emission \citep{LaDous_limb_darkening:1991A&A...252..100L,Matthews+2015MNRAS.450.3331M,Tomaru_winds_coronal_sources:2023MNRAS.523.3441T}.

In LMXBs, irradiation by the central X-ray source strongly influences the vertical temperature structure of the outer disc atmosphere \citep{KoKallman:1994ApJ...431..273K}, and the balance between net absorption and net emission depends sensitively on the hardness of the incident spectrum \citep{Sakhibullin_X-Ray_irr_opt_disks:1997AZh....74..432S,Shimanskii_opt_lines_irr_discs:2012ARep...56..741S}. In particular, softer irradiation deposits energy in shallower layers, favouring strong emission and weaker absorption troughs, whereas harder irradiation can deposit energy deeper in optically thick layers that contribute modestly to the line emission, but may allow absorption features to persist. This naturally explains the persistence of the broad (blue- and/or red-shifted) Balmer absorption troughs without the need to invoke both inflows and outflows.

For \src, the presence of systematic absorption troughs in both the blue and red wings, starting at similar velocities and with centroids consistent with zero (Section~\ref{sec: results}), supports a superposition of a broad absorption component with a narrower emission component. This interpretation is reinforced by the H$\alpha$ FWHM evolution across the hard--to--soft transition in Fig.~\ref{fig: line prop evo}. The H$\alpha$ FWHM exhibits the opposite behaviour to He{\sc ii} (which traces the ionised disc): when emission is embedded within absorption, a weakening of the emission core can produce an apparent decrease in the observed FWHM even if the intrinsic line-forming region is contracting, as suggested by the broadening of the He\,\textsc{ii} profiles. A similar effect is seen in the low-inclination nova-like variable V341\,Ara, where changes in the outer, photospheric disc affect the relative strength of the emission component \citep{CastroSegura2021MNRAS.501.1951C}.

\subsection{The disc's spectroscopic response to jet ejection(s)} \label{sec: jet-disc}

Epoch~7 coincides with the late-rising, near-peak luminosity of the brightest radio flare observed in \src\ \citep{Hughes2025ApJ...988..109H}. These flares are associated with the launch of discrete, bipolar ejecta during the hard-to-soft transition \citep{Hughes_LrLx:2025MNRAS.542.1803H}, and a reconfiguration of the inner accretion flow \citep[e.g. ][; see also \citealt{Fender_2004}]{Bright:2020NatAs...4..697B}. Figure~\ref{fig: line prop evo} suggests that the outer disc reacts to this event by reducing the peak-to-peak separation and FWHM, while simultaneously increasing the He\,\textsc{ii} flux. 
This is naturally explained by the extension of the chromospheric line-emitting region to larger radii, perhaps as a consequence of the outer disc becoming more strongly irradiated. This would naturally shift the emissivity to lower Keplerian velocities,  while boosting recombination lines from the X-ray heated atmosphere \citep[e.g. ][]{KoKallman:1994ApJ...431..273K,Dubus+2001AA...373..251D,Lasota2001NewAR..45..449L}. 
This interpretation is reinforced by the presence of an optically thick outflow during Epoch 7, as discussed in Section~\ref{sec: radio flaring}. As suggested by \citet{Tetarenko2018Natur.554...69T}, such an outflow can ``reflect'' radiation from the inner accretion flow, increasing the high-energy flux incident on the outer disc and enhancing its heating.
In the $\alpha$-prescription, the viscosity of the disc ($\nu$) at a given radius is proportional to the temperature $T$, with $\nu\propto \alpha T$. So irradiation-driven heating increases the viscosity and can even push the outer disc into the hot, ionized branch \citep{Dubus+2001AA...373..251D}. Consequently, the local mass accretion rate through the disc would increase, perhaps helping to drive the state transition. This effect may also introduce dynamic perturbations in the disc that might explain the misfit of simple models during the state transition observed in MAXI J1820+070 \citep{Georganti_maxi1820:2026MNRAS.545f1965G}. 
Notably, despite the pronounced outer–disc response at Epoch~7, we do not detect contemporaneous changes in the base of the double peaked emission lines \citep[a proxy for the inner–disc radius; ][]{BorisovNustroev_disc_lines:1997BSAO...44..110B}.

Overall, our data suggest that the jet-ejection episode drives a transient irradiation-induced redistribution of the optical line emissivity towards larger radii, while the innermost disc radius remains broadly unchanged within our sensitivity.

\subsection{A massive outflow in the dim-hard state} \label{sec: outflow discussion}

Perhaps the most striking feature in our data is the dramatic change triggered by the transition from the soft to the dim-hard state (Epochs 10 and 11). Here, the higher-order Balmer lines suddenly change from emission- to absorption-dominated, and a pronounced Balmer absorption edge. 
In these higher-order Balmer lines, the core of the absorption feature is filled with relatively weak emission. The relative depth and width of the absorption lines are sufficient to allow us to characterise their properties (see Section~\ref{sec: HTS results} and Table~\ref{tab: abs fit}. These measurements are consistent with the presence of a cold ($T\lesssim10^4\,\mathrm{K}$), optically thick, gas moving along the line of sight with a bulk velocity of $v_{w}\simeq -750\,\mathrm{km\,s^{-1}}$.

\begin{figure}
    \centering
    \includegraphics[width=0.98\linewidth]{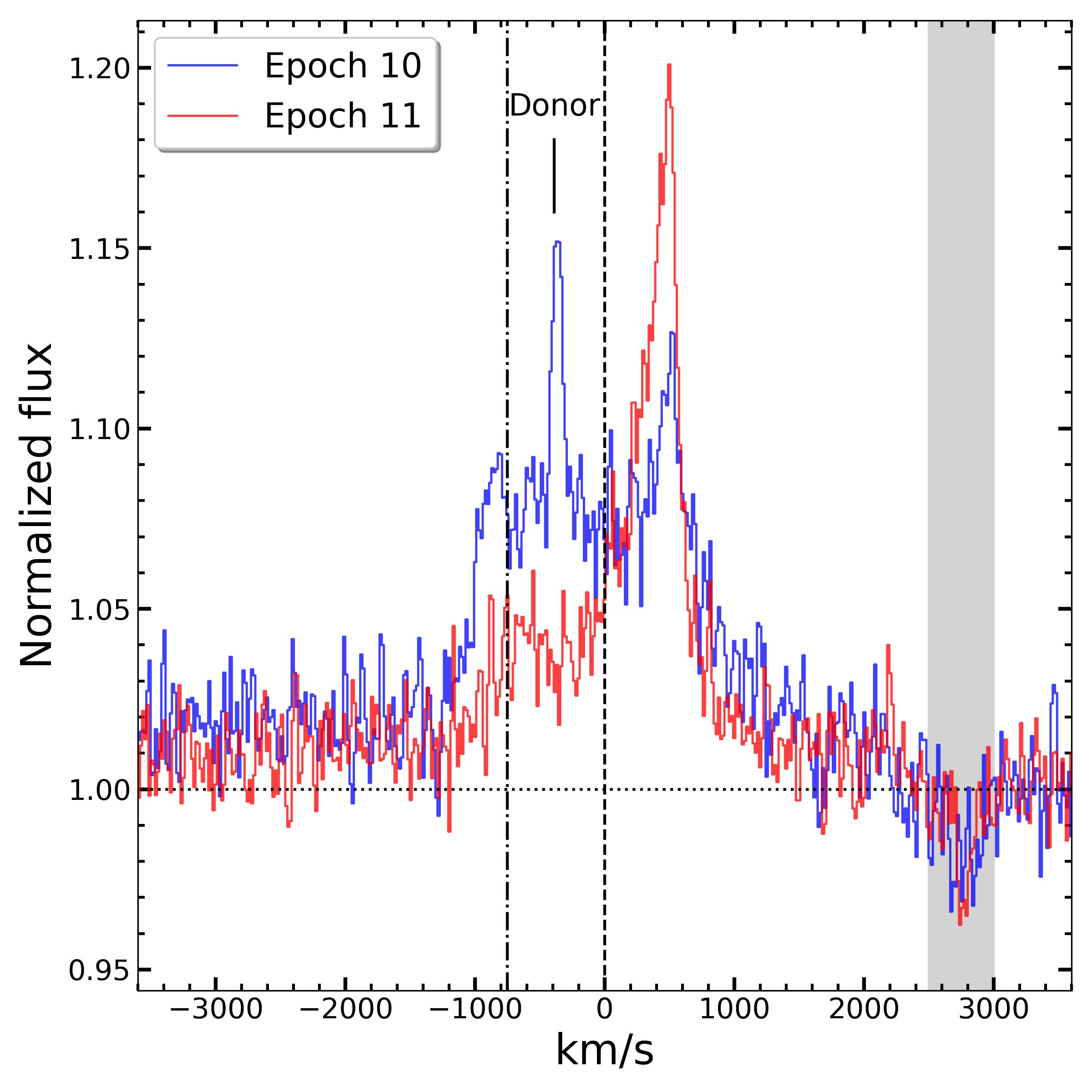}
    \caption{He\,\textsc{ii} $\lambda4686$ line profiles in Epochs 10 and 11, corresponding to the low-luminosity soft state and the low–hard state, respectively. In Epoch 10, the profile shows a bright narrow component at $v\simeq-400\,\mathrm{km\,s^{-1}}$, consistent with emission from the irradiated face of the donor, whose velocity during this epoch derived from the orbital solution is marked with the vertical tick labelled as "Donor".
    Epoch 11 is characterised by a broad base with an emission component on the red side; 
    on the blue side, a broad base extends to velocities similar to those in the previous epoch. 
    The strong, contemporaneous Balmer absorption at a estimated velocity $v_w\simeq-750\,\mathrm{km\,s^{-1}}$ (indicated with the vertical dot-dash line), suggests an underlying double-peaked disc line in which a foreground outflow absorbs the blue peak (c.f. Sec. \ref{sec: outflow discussion}). 
}
    \label{fig: epochs10-11}
\end{figure}

During this final epoch, He\,\textsc{ii} remains in emission but exhibits a characteristic shape. Given that the system is near orbital phase $\phi \approx 0$, the narrow component from the hotspot or companion is expected near the rest velocity ($v \simeq 0\,\mathrm{km\,s^{-1}}$); however, this feature is now negligible compared to Epoch 10. Instead, the line exhibits a weak, but noticeable, broad base with a single emission component on the red side. As can be seen in Figure~\ref{fig: epochs10-11}, the shapes of the red peak and base of the line are similar to those in the soft state, suggesting a disc-like, double-peaked profile where the blue component has been absorbed. 

Notably, the velocity of this 'missing' blue component roughly matches the blueshifted absorption observed in the Balmer lines, pointing toward a unified origin. We primarily interpret these broad, blueshifted features as the signature of a massive, cool accretion disc outflow. As discussed in Section~\ref{sec: balmer abs}, the hardening of the X-ray spectrum between Epochs 10 and 11 likely plays a crucial role here; by depositing energy deeper into the optically thick layers of the disc, this harder irradiation may facilitate the reappearance of a disc photosphere. Consequently, the broad Balmer absorption could represent a combination of both a photospheric contribution and an outflowing component. However, while the disc photosphere can account for the presence of absorption troughs, the striking kinematic alignment between the Balmer absorption and the He\,\textsc{ii} asymmetry suggests that a line-of-sight wind provides a more cohesive physical explanation for the specific profiles observed. This leads us to consider a unified model where the lines are shaped by the kinematics of the wind, which we explore in the following section.

\subsubsection{Line formation in an accelerating and rotating disc wind}

\begin{figure*}
    \centering
    \includegraphics[width=0.99\linewidth]{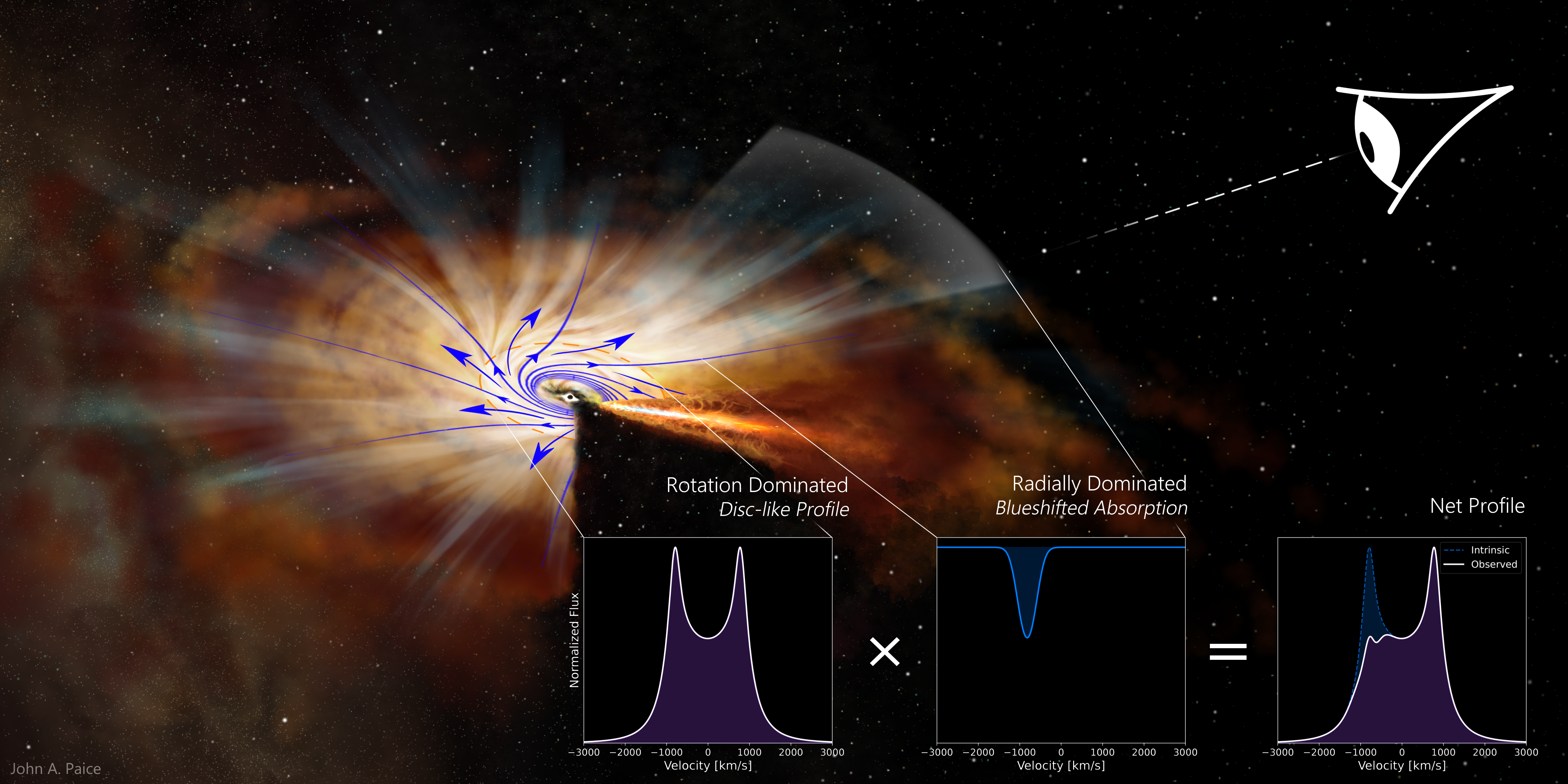}

    \caption{Schematic of an outflow producing double peaked emission lines in the (rotation dominated), inner regions and becomes radially dominated a larger radii producing a blue-shifted absorption profile on the same spectral line. Such dynamic would produce the single line emission with broad absorption shown in Figure~\ref{fig: epochs10-11}.}
    \label{fig: Schematics}
\end{figure*}

A natural way to interpret the Balmer and He\,\textsc{ii} line profiles seen in Epoch 11 is to assume that both sets of lines are formed in a rotating and accelerating outflow. The basic physical picture is illustrated in Figure~\ref{fig: Schematics}. The kinematic of the line-forming region are rotationally dominated close to the inner disk, but become outflow-dominated further out. Both the Balmer transitions and the He\,\textsc{ii} line are recombination features whose emissivity scales with density squared. The unabsorbed emission lines are therefore produced in the high-density base of the wind, with profiles that exhibit the classic double-peaked shape associated with lines formed in rotation-dominated flows.
By contrast, absorption scales with optical depth, which is linear in density. The blue-shifted absorption features in the Balmer lines, as well as the suppression of the blue peak of the He\,\textsc{ii} line, can therefore be produced further out in the wind, where the density is lower and the kinematics are outflow-dominated.

\begin{figure}
    \centering
    \includegraphics[width=0.95\linewidth]{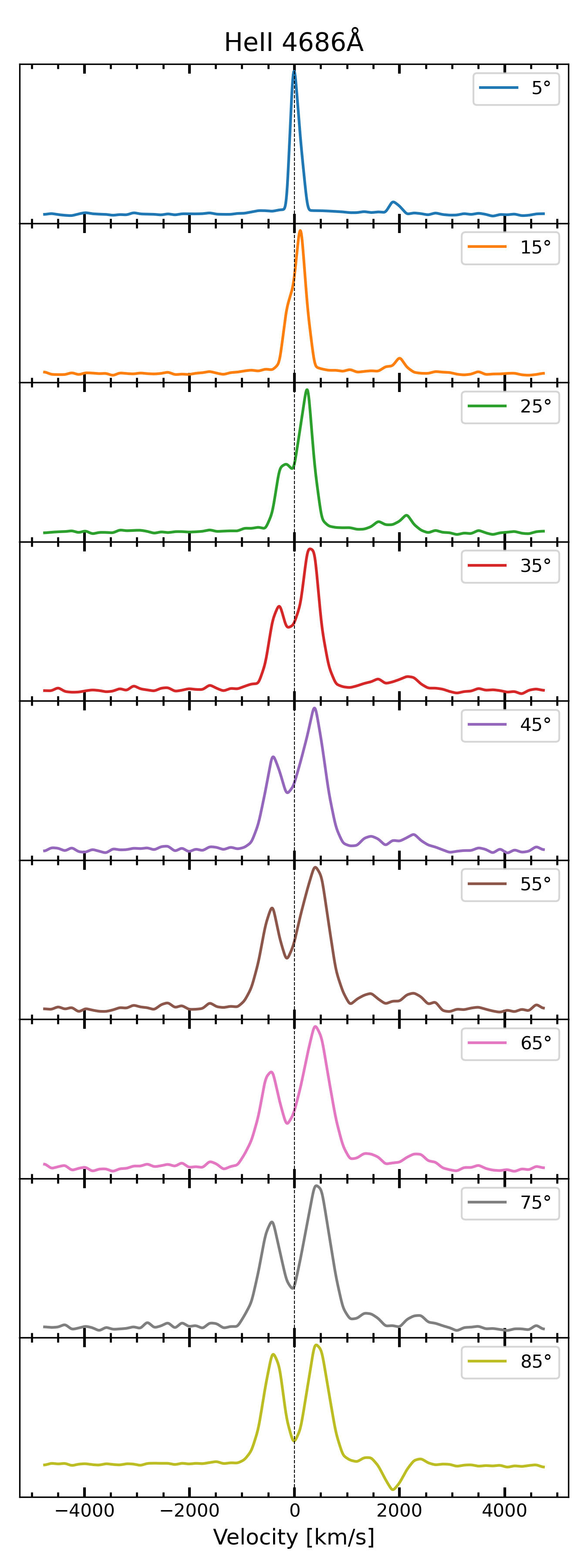}
    \caption{Synthetic spectrum around He\,\textsc{ii} $\lambda4686$ generated with \textsc{SIROCCO} using the model presented in \citet{Koljonen_Maxi1820:2023MNRAS.521.4190K} with hybrid macro-atom scheme \citep[c.f.][]{SIROCCO:2025MNRAS.536..879M}, and $\dot{M}_{w} = 1.4\times10^{-7}\mathrm{M_\odot\,yr^{-1}}$. The model produces line profiles resembling to those observed in Fig.~\ref{fig: epoch11 only} for a range of moderate inclinations. 
}
    \label{fig: sirocco}
\end{figure}

In order to test whether a rotating outflow can actually give rise to the unusual He\,\textsc{ii} line profile observed in Epoch 11, we also carried out a small number of exploratory radiative transfer simulations using \textsc{SIROCCO} \citep{LongKnigge2002ApJ...579..725L,SIROCCO:2025MNRAS.536..879M}. We focus here on a model similar to the default LMXB template used in \citet[][]{Koljonen_Maxi1820:2023MNRAS.521.4190K}. Such a model can produce reasonable optical emission line spectra provided the mass-loss rate is high enough -- and/or the filling factor low enough -- to create high enough densities near the wind base. As shown in Figure~\ref{fig: sirocco}, this model naturally produces double-peaked emission lines with (partially) suppressed blue peaks for a significant range of viewing angles. It would clearly be interesting and important to determine the exact conditions that produce this behaviour, but this is beyond the scope of the present study. Our key point here is simply that rotating and accelerating disc winds {\em can} produce the unusual line profile shapes we observe in Epoch 11.

\subsubsection{On the detectability of the outflow}

Optical disc winds in LMXBs are typically observed early in the luminous hard state and during the transition toward the soft state \citep[e.g.][and references therein]{winds_review:2026arXiv260105319M}. In the case of \src, we do not see unambiguous evidence of the smoking-gun signatures of disc winds (i.e. P-Cygni profiles). Instead, we argue that the emission during the luminous hard-to-soft state transition is likely dominated by a combination of chromospheric and photospheric components from the accretion disc (cf. Sec.~\ref{sec: balmer abs}). But does this imply a disc wind absent during this period? The short answer is no.
As shown in \citet{Castro-Segura2022Natur.603...52C}, the detectability of the lines can be diluted by changes in luminosity from a component whose emission does not pass through (or get reprocessed by) the outflow. So even if the outflow signatures are present in the spectrum, sufficiently luminous background emission can dilute them to the point of being unobservable. This argument matches the expectations of \src, where the inclination is thought to be moderate–low; therefore, the contribution from the inner disc and/or the base of the jet can easily outshine an outflow component, rendering it unobservable. Alternatively, the outflow may be over-ionised during the luminous states \citep[c.f. ][]{Munoz-Darias2016}.

\subsubsection{The wind mass loss rate}

We can use the Sobolev approximation \citep{Sobolev1960mes..book.....S} to roughly estimate the wind mass-loss rate that is required to produce the observed blue-shifted absorption in the higher-order Balmer lines. In any moving medium, a photon can only interact with a particular bound-bound transition if its frequency matches the Doppler-shifted rest frequency of the transition, $\nu_0$, to within a thermal Doppler width, $\Delta \nu_{\rm th}$. 

If the velocity gradient in the medium is large, interactions can happen in a small resonance region whose Sobolev width is $l_{S} \simeq \Delta v_{\rm th} / \left| dv/dl \right| $ \citep[e.g.][]{Castor1970MNRAS.149..111C}. Here, $dv/dl$ is the velocity gradient (measured along the photon's line-of-flight), and $\Delta v_{\rm th} = c (\Delta \nu_{\rm th}/\nu_0)$ is the thermal Doppler width in velocity units. Within this small region, the physical properties of the wind can therefore be assumed to be constant. 

The Sobolev optical depth the photon will encounter in crossing the resonance region is 
\begin{equation}\label{eq:tau_s}
\displaystyle
\tau_S \simeq
\frac{\pi e^2}{m_e c}\,
\frac{f_{\ell u}\,\lambda_0}{\left| dv/dl \right|}
\, n_\ell, 
\end{equation}
where we neglected the modest contribution from stimulated-emission  \citep{Castor1970MNRAS.149..111C,Rybicki_Hummer:1978ApJ...219..654R,Lamers_Cassinelli:1999isw..book.....L}.

Here, $f_{\ell u}$ is the f-value and $\lambda_0 = c/\nu_0 $  the rest wavelength of the relevant transition. For radially streaming photons in a spherical outflow, $dv/dl = dv/dr$, where $v(r)$ is the velocity law in the wind (i.e. the velocity as a function of radius). For any given $dv/dr$, Equation \ref{eq:tau_s} provides an estimate of the line optical depth at the (projected) velocity that corresponds to the assumed $dv/dr$ \citep[e.g.][]{Lamers:1987ApJ...314..726L}. 

During Epoch 11, the higher-order Balmer lines exhibit blue-shifted absorption features around $v \simeq -750\,\mathrm{km\,s^{-1}}$. Since this cold wind presumably emanates from the outer disc, let us adopt a characteristic length-scale of $\Delta l \sim 10^{10}$ cm and assume that it accelerates by $\Delta v \sim 750\,\mathrm{km\,s^{-1}}$ over this length scale. So, along the photon's path, we can then adopt $dv/dl \simeq \Delta v / \Delta l$. 

Substituting this into Equation \ref{eq:tau_s}, along with the f-value and rest wavelength appropriate for H$\gamma$, gives a simple scaling relation for the Sobolev optical depth in this line
$$
\tau_S(\mathrm{H}\gamma)
\approx
0.7 \,
\left(\frac{n_2}{10^{5}\, \mathrm{cm^{-3}}}\right)
\left(\frac{\Delta l}{10^{10}\, \mathrm{cm}}\right)
\left(\frac{\Delta v}{750\, \mathrm{km\,s^{-1}}}\right)^{-1}.
$$
In order to produce the observed absorption feature, we require $\tau_S \sim 1$, and therefore $n_{2} \simeq 10^{5}\, \mathrm{cm^{-3}}$.

In the spirit of the order-of-magnitude estimate we are interested in here, we now assume LTE conditions to obtain a rough upper limit on the fraction of Hydrogen atoms or ions in the $n=2$ level of HI. Under these conditions, it can be shown that 
$$\frac{n_2}{n_H}\bigg|_{\rm max} \simeq 8 \times 10^{-12}  \left(\frac{n_H}{10^6 {\rm cm}^{-3}}\right)^{0.71},$$
where the subscript ``max'' indicates that this is the maximum ratio that can be obtained regardless of temperature. In practice, this maximum is found at physically reasonable temperatures near $T \simeq 10^4$K. If we conservatively adopt this maximum fraction to estimate the required $n_2$, we obtain
$$n_2 \simeq 8 \times 10^{-6}  \left(\frac{n_H}{10^6 cm^{-3}}\right)^{1.71}\quad\mathrm{cm^{-3}}.$$
We can therefore estimate the total Hydrogen density required in the line-forming region
$$
n_H \;\approx\; 8 \times10^{11}\,
\left(\frac{n_2}{10^{5}\ \mathrm{cm^{-3}}}\right)^{0.585}
\quad\mathrm{cm^{-3}}.
$$
As expected, this is quite a high density. The formation of Balmer absorption features, even under favourable conditions, requires high densities. It therefore also requires high mass-loss rates. If we approximate the outflow as quasi-spherical, we can use the continuity equation to link the mass-loss rate, $\dot{M}_w$ to the Hydrogen number density via 
$$
\dot{M}_w = 4\pi R^2 \mu m_p\, n_H(R)\, v(R),
$$
where $\mu \simeq 0.6$ is the mean molecular weight per proton. If we once again adopt characteristic values of $R \simeq 10^{10}\,\mathrm{cm}$ and $v \simeq -750\,\mathrm{km\,s^{-1}}$, we obtain
$$
\dot{M}_w \;\approx\;
1 \times10^{-9}\ M_\odot\,\mathrm{yr^{-1}}\,
\left(\frac{n_H}{8\times10^{11}\ \mathrm{cm^{-3}}}\right)
\left(\frac{r}{10^{10}\ \mathrm{cm}}\right)^{2}
\left(\frac{v}{750\ \mathrm{km\,s^{-1}}}\right).
$$

This estimate is obviously approximate. First, the assumption of LTE is clearly an oversimplification. On the other hand, at the high densities and opacities required to produce the observed features, LTE may actually be quite a reasonable approximation. Also, by adopting the {\em maximum} population of $n_2$ (i.e. regardless of temperature), our estimates of $n_H$ and $\dot{M}_w$ should be fairly conservative, i.e. subject to the LTE assumption, they are lower limits on the true values.

Second, the geometry of the wind is clearly uncertain: a disc wind subtends a solid angle $\Omega<4\pi$, so $\dot{M}_w$ should strictly be scaled by $\Omega/4\pi$. However, this ratio is likely to be $O(1)$, since disc wind signatures have been observed in systems spanning a reasonably wide range of inclinations. Third, clumping or partial covering may complicate the mapping between the observed absorption depth, $\tau_S$, $n_H$ and $\dot{M}_w$  \citep[e.g. ][]{Koljonen_Maxi1820:2023MNRAS.521.4190K}. Fourth and finally, the characteristic radius and kinematics are highly uncertain. Since  $\tau_S \propto n_2 \left| dv/dl \right|$ and $\dot{M}_w \propto R^2n_Hv$, these uncertainties alone imply that our inferred $\dot{M}_w$ should be viewed as an order-of-magnitude estimate (or perhaps lower limit), rather than a precise estimate.

\subsubsection{Implications} \label{sec:wind_implications}

 The order-of-magnitude estimates above suggest that producing Balmer absorption at $v\simeq-750~\mathrm{km\,s^{-1}}$ requires a dense outflow, with a characteristic mass-loss rate of $\dot{M}_w \sim 10^{-9}\,M_\odot\,\mathrm{yr^{-1}}$ under our adopted assumptions (and potentially larger if the $n=2$ population is lower than the maximum LTE value or if the characteristic radius is larger). 
This allows us to assess whether such an optical wind could be dynamically important. 
X-ray spectral modelling of contemporaneous observations of \src\ during Epochs~10 and 11 yields an inner-disc accretion rate of $\dot{M}_{\rm acc}\simeq10^{-9}\,(D/3.5\,\mathrm{kpc})^2\,M_\odot\,\mathrm{yr^{-1}}$ (Brigitte et al.\ in prep.).
The inferred $\dot{M}_w$ is therefore at least a substantial fraction of $\dot{M}_{\rm acc}$ during the dim-hard state (i.e. $\dot{M}_w/\dot{M}_{\rm acc}\sim1$ for the parameters adopted here).
If disc winds persist throughout the outburst as suggested by previous studies \citep{SanchezSierras2020AA...640L...3S,Castro-Segura2022Natur.603...52C,CastroSegura2023}, this naturally points towards strongly non-conservative mass transfer and potentially enhanced angular-momentum losses, with direct implications for the accretion/ejection balance during outburst \citep[e.g.][]{Tetarenko2018MNRAS.480....2T}.

\src\ is a transient BH LMXB, and the disc-instability framework constrains the average mass-transfer rate from the companion $\langle \dot{M}_{\rm tr} \rangle$ to $\lesssim10^{-8}\,M_\odot\,\mathrm{yr^{-1}}$ \citep[e.g.][]{Coriat+2012MNRAS.424.1991C,Dubus2019A&A...632A..40D}. If such dense outflows were confirmed to operate not only in outburst but also down to X-ray luminosities comparable to quiescent levels, an outflow at the level inferred here would imply that at least $\sim10$\% of the mass being transferred is continuously being blown away. In that case, disc winds could have an even stronger impact on the secular evolution of BH LMXBs than previously thought, by enforcing significantly non-conservative mass transfer on evolutionary timescales \citep[e.g.][]{Gallegos-Garcia:2024ApJ...973..168G,Fijma:2025MNRAS.544.4702F}.

\section{Conclusions}\label{sec: CCL}
In this work, we presented state-resolved spectroscopy of the black hole low-mass X-ray binary \srclong\ during its 2023 discovery outburst gathered with VLT/X-Shooter. \src\ followed the canonical q-shaped track, sampling the luminous hard state (epochs 1–5), the luminous soft state (epochs 6–9), and the late soft-to-hard transition (epochs 10-11).

During the luminous hard and soft states, the optical spectrum is dominated by disc-like emission lines. He\,\textsc{ii} exhibits persistent double-peaked profiles consistent with rotationally dominated (disc) origin, while the Balmer series also show frequent transient absorption troughs in the wings. 
These troughs often appear systematically in the red and blue wings and, when measurable, remain consistent with being centred at the rest wavelength, favouring a photosphere+chromosphere/wind superposition whose balance is regulated by irradiation. 

We find evidence for a transient response of the outer disc during jet launching. Epoch 7, obtained near the peak luminosity of the brightest radio flare, shows a small but significant increase in He\,\textsc{ii} flux accompanied by a decrease in peak-to-peak separation and FWHM. This behaviour is naturally explained if enhanced irradiation redistributes the optical line emissivity towards larger radii, reducing the characteristic Keplerian velocities while boosting recombination emission from the heated disc atmosphere. Notably, we do not detect contemporaneous changes in the base of the double-peaked profiles, suggesting that any inner-disc response is below our sensitivity owing to the small radius of this region.

The transition to the dim-hard state in Epoch 11 is marked by the emergence of broad Balmer absorption and a characteristic asymmetry in the He\,\textsc{ii} emission. We suggest that the hardening of the incident X-ray spectrum may contribute to Balmer absorption features by allowing a disc photosphere to persist; however, the measured blueshift of $v_w\simeq-750\,\mathrm{km\,s^{-1}}$ and the simultaneous suppression of the He\,\textsc{ii} blue peak point toward a dynamic origin. We therefore interpret these profiles as a combination of a photospheric contribution and a massive, cool disc outflow. A comparison with \textsc{SIROCCO} radiative transfer simulations demonstrates that a rotating and accelerating wind can naturally reproduce these morphologies. In this framework, the double-peaked emission is formed in the high-density, rotationally-dominated inner outflow, while the blueshifted absorption features arise further out where the kinematics are dominated by radial expansion.

Finally, we obtained an order-of-magnitude lower limit on the wind mass-loss rate using the Sobolev formulation, finding $\dot{M}_w \gtrsim 10^{-9}\,M_\odot\,\mathrm{yr^{-1}}$ for our adopted parameters. 
This lower limit is comparable to the instantaneous mass accretion rate derived from X-ray observations ($\dot{M}_w/\dot{M}_{\rm acc}\sim1$) and about 10 per cent of the average mass transfer rate from the companion star. 
  
Taken together, these results support a picture in which irradiation governs the balance between photospheric absorption and chromospheric/wind emission during the bright state, in line with \citet{Castro-Segura2022Natur.603...52C}. They also show that jet-launching episodes can drive rapid, measurable changes in the outer-disc line-emissivity profile, and that a dense, cool, optically thick outflow can emerge (or become detectable) at late times even when the X-ray luminosity is surprisingly low.

If such dense outflows were confirmed to operate not only in outburst but also down to X-ray luminosities comparable to quiescent levels, an outflow like the one inferred here would imply that at least $\sim10$ per cent of the transferred mass is continuously expelled. In that case, disc winds could play a more important role in the secular evolution of BH LMXBs than previously thought, by enforcing significantly non-conservative mass transfer on evolutionary timescales.

\section*{Acknowledgements}

NCS and DLC acknowledge support from the Science and Technology Facilities Council (STFC) grant ST/X001121/1. IP acknowledges support from the Royal Society through a University Research Fellowship (URF\textbackslash R1\textbackslash 231496). DS acknowledges support from the Science and Technology Facilities Council (STFC), grant numbers ST/T007184/1, ST/T003103/1, ST/T000406/1, ST/X001121/1 and ST/Z000165/1. Partial support for KSL's effort on the project was provided by NASA through grant numbers HST-GO-16489 and HST-GO-16659 from the Space Telescope Science Institute, which is operated by AURA, Inc., under NASA contract NAS 5-26555. AJT acknowledges that this research was undertaken thanks to funding from the Canada Research Chairs Program and the support of the Natural Sciences and Engineering Research Council of Canada (NSERC; funding reference number RGPIN--2024--04458).

Based on Director's Discretionary Time observations collected at the European Southern Observatory under ESO programmes 111.265Y and 112.26W7. We thank the ESO staff for their rapid response and flexible scheduling of these programs. 
This work has made use of the Pan-STARRS1 survey. The Pan-STARRS1 Surveys (PS1) and the PS1 public science archive have been made possible through contributions by the Institute for Astronomy, the University of Hawaii, the Pan-STARRS Project Office, the Max-Planck Society and its participating institutes, the Max Planck Institute for Astronomy, Heidelberg and the Max Planck Institute for Extraterrestrial Physics, Garching, The Johns Hopkins University, Durham University, the University of Edinburgh, the Queen's University Belfast, the Harvard-Smithsonian Center for Astrophysics, the Las Cumbres Observatory Global Telescope Network Incorporated, the National Central University of Taiwan, the Space Telescope Science Institute, the National Aeronautics and Space Administration under Grant No. NNX08AR22G issued through the Planetary Science Division of the NASA Science Mission Directorate, the National Science Foundation Grant No. AST-1238877, the University of Maryland, Eotvos Lorand University (ELTE), the Los Alamos National Laboratory, and the Gordon and Betty Moore Foundation.  
PAC acknowledges the Leverhulme Trust for an Emeritus Fellowship.

\section*{Affiliations}
\noindent
{\it
$^{1}$Department of Physics, University of Warwick, Gibbet Hill Road, Coventry CV4 7AL, UK\\
$^{2}$Department of Astronomy, University of Cape Town, Private Bag X3, Rondebosch 7701, South Africa\\
$^{3}$South African Astronomical Observatory, PO Box 9, Observatory 7935, Cape Town, South Africa\\
$^{4}$European Southern Observatory, Alonso de Córdova 2107, Vitacura, Casilla 19001 Santiago de Chile, Chile\\
$^{5}$Department of Physics and Astronomy, University of Southampton, Southampton, Hampshire, SO17 1BJ, United Kingdom\\
$^{6}$Department of Physics, University of the Free State, 205 Nelson Mandela Drive, Bloemfontein, 9300, South Africa\\
$^{7}$Astrophysics, Department of Physics, University of Oxford, Keble Road, Oxford OX1 3RH, UK\\
$^{8}$Astronomical Institute of the Czech Academy of Sciences, Bo\v{c}n\'{i} II 1401, CZ-14100 Praha 4, Czech Republic\\
$^{9}$Astronomical Institute, Charles University, V Hole\v{s}ovi\v{c}k\'{a}ch 2, CZ-180 00 Prague 8, Czech Republic\\
$^{10}$European Southern Observatory (ESO), Karl-Schwarzschild-Strasse 2, 85748 Garching bei M\"{u}nchen, Germany\\
$^{11}$INAF -- Osservatorio Astronomico di Roma, Via Frascati 33, I-00078 Monte Porzio Catone (RM), Italy\\
$^{12}$Instituto de Astrof\'{i}sica de Andaluc\'{i}a (IAA-CSIC), Glorieta de la Astronom\'{i}a s/n, E-18008, Granada, Spain\\
$^{13}$Space Telescope Science Institute, 3700 San Martin Drive, Baltimore, MD 21218, USA\\
$^{14}$Eureka Scientific, Inc., 2452 Delmer Street Suite 100, Oakland, CA 94602-3017, USA\\
$^{15}$INAF -- Istituto di Astrofisica Spaziale e Fisica Cosmica di Palermo, Via Ugo La Malfa 153, I-90146 Palermo, Italy\\
$^{16}$Department of Physics and Astronomy, University of Lethbridge, Lethbridge, Alberta, T1K 3M4, Canada\\
$^{17}$Centre for Fluid and Complex Systems, Coventry University, Coventry CV1 5FB, UK
}
\section*{Data Availability}
The data underlying this article is publicly available in the corresponding telescope archives: \url{https://archive.eso.org/} for VLT/X-Shooter and \url{https://ssda.saao.ac.za} for SALT. 
 



\bibliographystyle{mnras}



\newpage
\appendix
\section{Full Observing log}
\
\begin{table*}
\centering
\caption{Summary of X-Shooter observations of Swift J1727.8--1613 with associated orbital phase, observation date, accretion state, number of exposures, and exposure times for each arm. Epochs 9 and 10 each have two observations. For Epoch 11, the NIR observations consist of 12 exposures, compared to only 4 in the UVB and VIS arms. }
\begin{tabular}{cccccccr}
\toprule
\quad \textbf{Epoch} & \textbf{Orbital} & \textbf{Date} & \textbf{MJD} & \textbf{Accretion State} & \textbf{N Exp} & \textbf{Exposure Time [s]}  \\
 & \textbf{phase} & &  & &  & (UVB/VIS/NIR) \\ 
\midrule
\multicolumn{1}{l}{Hard-to-soft}\\
\midrule
\quad 1   & 0.76 & 2023-08-27 & 60183.01 & Hard & 12 & 131 / 120 / 149  \\
\quad 2   & 0.62 & 2023-09-04 & 60191.05 & Hard & 16 & 58 / 47 / 64  \\
\quad 3   & 0.79 & 2023-09-14 & 60201.03 & Hard & 4 & 108 / 97 / 112  \\
\quad 4   & 0.40 & 2023-09-26 & 60213.01 & Hard & 4 & 118 / 107 / 120  \\
\quad 5   & 0.65 & 2023-10-01 & 60218.98 & Hard & 4 & 167 / 122 / 200   \\
\quad 6   & 0.01 & 2023-10-07 & 60224.99 & Soft & 6 & 167 / 122 / 200  \\
\quad 7   & 0.39 & 2023-10-14 & 60231.01 & Soft & 4 & 167 / 122 / 200  \\
\quad 8   & 0.04 & 2023-10-17 & 60234.01 & Soft & 4 & 167 / 122 / 200 \\
\quad 9(2)   & 0.14 & 2023-10-21 & 60239.00 & Soft & 8 & 167 / 122 / 200 \\
\midrule
\multicolumn{1}{l}{Soft-to-hard}\\
\midrule
\quad 10(2)  & 0.75 & 2024-02-29 & 60369.38 & Soft & 4 & 310 / 300 / 300 \\
\quad 11  & 0.02 & 2024-03-24 & 60393.36 & Hard & 4(12) & 509 / 519 / 175 \\
\bottomrule
\end{tabular}
\label{tab:xs_obs_full}
\end{table*}


\bsp	
\label{lastpage}
\end{document}